# Assessing Machine Learning Algorithms for Near-Real Time Bus Ridership Prediction During Extreme Weather

Francisco Rowe[1,*], Michael Mahony[1], Sui Tao[2]

[1]*Department of Geography and Planning, University of Liverpool, Liverpool, UK*
[2]*Faculty of Geographical Science, Beijing Normal University, Beijing, China*
*Corresponding author: F.Rowe-Gonzalez@liverpool.ac.uk

**Abstract**

Given an increasingly volatile climate, the relationship between weather and transit ridership has drawn increasing interest. However, challenges stemming from spatio-temporal dependency and non-stationarity have not been fully addressed in modelling and predicting transit ridership under the influence of weather conditions especially with the traditional statistical approaches. Drawing on three-month smart card data in Brisbane, Australia, this research adopts and assesses a suite of machine-learning algorithms, i.e., random forest, eXtreme Gradient Boosting (XGBoost) and Tweedie XGBoost, to model and predict near real-time bus ridership in relation to sudden change of weather conditions. The study confirms that there indeed exists a significant level of spatio-temporal variability of weather-ridership relationship, which produces equally dynamic patterns of prediction errors. Further comparison of model performance suggests that Tweedie XGBoost outperforms the other two machine-learning algorithms in generating overall more accurate prediction outcomes in space and time. Future research may advance the current study by drawing on larger data sets and applying more advanced machine and deep-learning approaches to provide more enhanced evidence for real-time operation of transit systems.

**Introduction**

Global warming resulting from the carbon emissions from human activities has become one of the most critical challenges for humanity (Nordhaus, 2007; Moser, 2010). Human-induced warming has raised the global temperature by approximately 1°C compared to the pre-industrial era according to a report by the Intergovernmental Panel on Climate Change (IPCC) (Allen *et al.*, 2018). The occurrence of extreme weather events, including extremely high and low temperatures, storms and hurricanes, has also increased in frequency since the 1950s (Stott, 2016). Such an increasingly versatile climate will considerably impact the natural and social environment, as such, potentially causing new issues and challenges for human societies (Urry, 2015; Allen *et al.*, 2018).

In combating global warming, interventions to curb carbon emissions are critical (Nordhaus, 2007; Zhang and Cheng, 2009). Particular attention has been drawn to the transportation sector (Loo and Li, 2012; Metz, 2015). The global collective contribution of road and other types of transport to carbon emissions has been estimated to be over 14% (IPCC, 2014). Promoting greener and attractive public transit service while reducing excessive car usage has been stressed as a key strategy (Friman *et al.*, 2013; Tao *et al.*, 2019). However, the increasing variability of weather associated with climate change may cause disruptive effects on the operation and management of urban transit systems (Hofmann and O'Mahony, 2005; Miao *et al.*, 2019), as well as demand for transit service (Böcker *et al.*, 2013; Tao *et al.*, 2018). How to ensure reliable and adequate transit service in response to sudden change of weather has attracted more concerns (Arana *et al.*, 2014; Singhal *et al.*, 2014).

A scrutiny of the transport literature reveals that real-time relationships between weather and public transport ridership has received increasing scholarly attention (e.g. Singhal *et al.*, 2014; Wei *et al.*, 2019). However, limited empirical research has simultaneously investigated the spatial and temporal dynamics of transit use in relation to weather conditions, which may restrain our ability to better predict transit demand and inform real-time operation of transit service. These deficits can be partly traced to the lack of timely longitudinal data public transport use at high spatial resolution and temporal resolutions (Tao *et al.*, 2018). Traditionally, time-use surveys and household travel diaries are used to study urban travel behaviour (Kitamura, 1988; Schlich and Axhausen, 2003). While valuable, these data sources are expensive, infrequent, offer low population coverage and become available with some delay following their gathering. Moreover, the spatial and temporal scales of such data are often limited due to high collection cost (Noland and Polak, 2002). New forms of data (or big data), such as smart card and mobile phone data have emerged as novel and useful sources to study human mobility. These data are generated at unprecedented temporal frequency and geographical granularity, enabling the analysis of human travel activity in real and near-real time scales.

Another key constraint to our current understanding of real-time relationships between weather and public transport ridership is a dearth in the application of simple, affordable, rapidly customisable and accurate predictive models. Such models could be key to bridge empirical research and practice in the operation of transit service. Predictive machine learning approaches enabled by affordable computing power emerge as a promising candidate. Unlike traditional statistical modelling methods, machine learning approaches do not necessarily rely on a predefined manual model specification (James *et al.*, 2013), and require limited human interventions in specifying the most appropriate model configuration for modelling urban traffic data associated with spatial and temporal dependence and non-stationarity (e.g. Zhou *et al.*, 2021). Recently, more transport studies have opted for machine learning approaches to examine and model transport phenomena (e.g., traffic flows and speed) in a timely and efficient manner (Ma *et al.*, 2015; Polson and Sokolov, 2017; Wang *et al.*, 2018). However, the application of machine learning approach to model and predict transit ridership vis-à-vis weather conditions has been relatively limited.

This article aims to address the identified gaps by using off-the-shelf machine learning models to produce near-real time, stop-level bus ridership predictions in response to sudden changes in weather conditions. In particular, we seek to:
(1) Model and predict the effects of current weather conditions on future patterns of bus ridership;
(2) Assess the accuracy of machine learning approaches in predicting future patterns of bus ridership under varying weather conditions; and,
(3) Generate implications for public transport planning and operation in response to extreme weather readings.

To these ends, we draw on a three-month smart card dataset of bus ridership, containing over 10 million observations allied with detailed weather measurements, trip length, calendar events, and built environment features to form an integrated spatio-temporal database. We train and assess three off-the-shelf tree-based machine learning algorithms, Random Forest (RF), eXtreme Gradient Boosting (XGBoost) and Tweedie XGBoost in their ability of short-term ridership prediction. We in particular assess the predictive model performance during periods of extreme temperature, rainfall, wind and humidity records. Tree-based machine learning algorithms provide an appropriate balance between interpretability and model complexity. Hence, they are preferred over more flexible but complex machine learning approaches, such as neural networks.

The rest of the paper is structured as follows: Section 2 provides a brief review of the relevant literature. Section 3 introduces the study context and data employed. Section 4 explains the machine learning approach developed and adopted in the current research. Section 5 presents and assesses the modelling and prediction results. Last, Section 6 summarises and discusses the main outcomes before a series of conclusions are drawn in Section 7.

## 2. Background

2.1. Impact of Weather on Transit Ridership

There is now wide recognition that people's daily travel arrangements are weather dependant (Böcker *et al.*, 2013; Liu *et al.*, 2017). Empirical evidence indicates that calm and pleasant weather tends to increase out-of-home travel, while unpleasant weather (e.g., cold and rainy periods) can exert the opposite effect (e.g. Cools *et al.*, 2010; Böcker *et al.*, 2013). Yet, the effects of weather on travel behaviour also tend to vary across travel of different purposes transport means and social groups (Palma and Rochat, 1992). Trips of obligatory nature (e.g., commuting, school trip) are often less affected by adverse weather conditions than trips of non-obligatory (e.g., leisure trips) (Sabir *et al.*, 2010). In terms of modal use, active transportation (i.e., cycling and walking) appears to be more elastic in response to weather than motorised modes (private and public transport) (Aaheim and Hauge, 2005; Liu *et al.*, 2015).

Increasing attention has also been given to the effect of weather on influencing public transit usage (Changnon, 1996; Hofmann and O'Mahony, 2005). Adverse weather conditions (e.g., heavy snow and rain) tend to influence people's decision of using public transport through impacting service quality (e.g., reliability, frequency) (Hofmann and O'Mahony, 2005) as well as individual travel experience (Miao *et al.*, 2019). A growing number of studies have probed into the daily variation of transit ridership vis-à-vis weather conditions. For example, a Chicago-based study showed that heavy rain, snow and strong wind significantly reduced bus and to a lesser extent, metro ridership (Guo *et al.*, 2007). Stover and McCormack (2012) confirmed that rainfall had a significant effect on influencing daily bus ridership in Washington city. (Kalkstein *et al.*, 2009) found an increase of transit ridership in association with dry, comfortable weather, and a decrease of ridership under cold, moist periods across three US metropolitan areas. In Brisbane, Australia, Kashfi (2016) found that seasonal changes and calendar events (e.g. university semester breaks) were dominant influencing factors of bus ridership, while the effects of weather conditions were comparatively less influential.

More recent studies have heeded weather-transit ridership at finer spatial and temporal granularities, enabled by the availability of smart card data that store detailed time and location information of transit trips, which provided more nuanced insights compared to the previous research. For example, using smart card data in Shenzhen (China), Zhou et al. (2017) modelled the influence of weather on the hourly metro service usage across system, stop and individual levels. They found that metro and bus ridership responded differently to weather variables, such as humidity and wind. In Brisbane, Australia, Tao *et al.* (2018) modelled hourly bus ridership at the stop level using time-series models over a three-month period, highlighting certain spatial heterogeneity in the response of bus usage across stops characterised by different built environment and land use patterns. In New York, Najafabadi et al. (2019) found that metro ridership within the commercial zones was less sensitive to the influence of rainfall than those in the residential zones especially during weekdays. In the Salt Lake City region, Miao et al. (2019) found that bus shelter helped retain or even increase stop-level daily ridership not only during heavy rains, but also days with extremely high and low temperatures.

The accumulating literature is suggestive that weather can cause significant short-term fluctuations in public transport usage, which deserve particular attention in the operation of public transport systems (Guo *et al.*, 2007; Tao *et al.*, 2018). Short-term transit ridership prediction that explicitly accounts for weather emerges as a promising strategy, as it can help inform adjustment to local transit services in response to sudden change in demand (Najafabadi *et al.*, 2019). However, few have tapped on this dimension. This may, in part, be hindered by the complexity of constructing the most-fitted statistical model to reliably capture the spatial-temporal relationship between weather and transit ridership. Advanced modelling methods to capture complex space-time interactions have been developed (Cheng *et al.*, 2014; Fotheringham *et al.*, 2015; Najafabadi *et al.*, 2019). Yet, methods capable of handling the spatial and temporal properties of weather and transit ridership and their interactions in an automatic manner are arguably still desirable.

2.2. Use of machine learning for bus ridership prediction

Recently, the application of machine learning has seen increasing interest in the arena of urban transportation research (Nguyen *et al.*, 2018; Xie *et al.*, 2020). Machine learning can be considered as a family of non-parametric modelling techniques developed and popularised in the field of computational and artificial intelligence (Jordan and Mitchell, 2015). Compared to the conventional statistical methods, machine learning techniques have more relaxed (or even absent) pre-assumptions about data distribution, and therefore are characterised by stronger computation capability especially in handling data of complex patterns and large quantities (Najafabadi *et al.*, 2015; Liu *et al.*, 2019). As such, machine-learning approaches appear to be suitable for traffic demand modelling, traffic signal control and various other prediction applications, which are often associated with strong spatial and temporal dependencies and heterogeneity (Baek and and Sohn, 2016; Nguyen *et al.*, 2018). In parallel, the emergence of big and open data in transport research has also rendered machine learning a suited analytical toolkit (Chen *et al.*, 2016; Wang *et al.*, 2018).

A main area of application of machine-learning approaches in transport research pertains to traffic flow simulation and prediction (Jin and Sun, 2008; Lippi *et al.*, 2013; Ma *et al.*, 2015; Nguyen *et al.*, 2018). Given the irregular and unpredictable behaviours in daily traffic (Liu *et al.*, 2019; Xie *et al.*, 2020), scholars have adopted machine learning to better capture the dynamics in traffic flows and make more reliable predictions. For example, using machines learning models, Lippi et al. (2013) simulated and predicted short-term traffic flow across the road network in California, which handled time-series data with reasonable efficiency. In Beijing, Ma et al. (2015) proposed a neural network model and applied to microwave sensor data collected along an expressway, which has shown enhanced capabilities in addressing time-series traffic issues. Chen (2017) also constructed a neural network model aimed particularly for big-data analysis. In Chicago, Polson and Sokolov (2017) developed a machine-learning architecture to improve traffic flow predictions under non-recurrent conditions (e.g., sport game, snowstorm).

Concerning public transit, growing interest has revolved around the potential of machine learning for modelling and predicting passenger flows and demand especially with smart card data (Sun *et al.*, 2015; Kurauchi and Schmöcker, 2017; Zhao *et al.*, 2018). Yet, empirical studies assess machine-learning approaches in predicting public transit usage has remained relatively limited (Nguyen *et al.*, 2018). For example, Baek and Sohn (2016) developed a neural network model to forecast bus ridership at the stop and stop-to-stop segment levels in Seoul, South Korea. Due to the large number of input features (over 9,000) coupled with a small sample of observations (five weekdays), the prediction error at the stop level was relatively low (<50%), while higher at the segment level (>70%). Liu et al. (2019) presented a deep learning neural-network algorithm to predict short-term metro passenger flows. Their results highlight that in addition to periodic patterns and trends of passenger behaviour, incorporating domain knowledge of metro systems also contributes to higher prediction accuracy. Also focusing on metro systems, Wei and Chen (2012) combined empirical mode decomposition with neural networks, which resulted in satisfactory prediction outcomes in terms of Mean Absolute Percent Error (MAPE) (between 6-10%), compared to SARIMA models (around 30%).

Despite the growing interest in applying machine learning in traffic flow and demand prediction, there remains considerable room for expanding machine-learning based research focusing on public transit. As noted above, while weather has been shown to have significant impact on transit ridership, few have employed machine learning to embark on short-term transit ridership prediction in relation to weather, and how it may inform transit operation in light of an increasingly volatile climatic context. Given that both weather and transit ridership are associated with strong spatial and temporal dependencies (Kashfi *et al.*, 2016; Najafabadi *et al.*, 2019), machine learning may serve as a suited tool for untangling and better capturing their complex interrelationship. This study will seek to partially address this gap by investigating and predicting bus ridership vis-à-vis near real-time weather conditions.

## 3. Study Context and Data

### 3.1. Study Context

Brisbane, the capital of the Queensland state, is the study context (Fig. 1). It is located at the east coast of Australia, and the third largest Australian city, with a population of over 2 million across the greater metropolitan area and 1 million within the city council area. In this study, we focus on the latter. As with other car-dependent cities, over 85% of all daily trips are by private vehicles (mainly cars), while the remaining are taken by public transport and active transport (BITRE, 2014). Brisbane's public transport system mainly consists of train, bus and ferry. Comprising over 400 routes, bus transit plays a key role in accommodating intra-city transit trips (Tao *et al.*, 2014). The construction of several busway segments since the early 2000 has been the major investment in Brisbane's bus transit system. The busway networks have been serving exclusively for the bus services connecting the city centre with the north, south and east suburbs of Brisbane, which collectively account for over half of the bus ridership in the city (Tao *et al.*, 2014). Compared to other on-road bus stops, busway stations also provide better shelters against the weather.

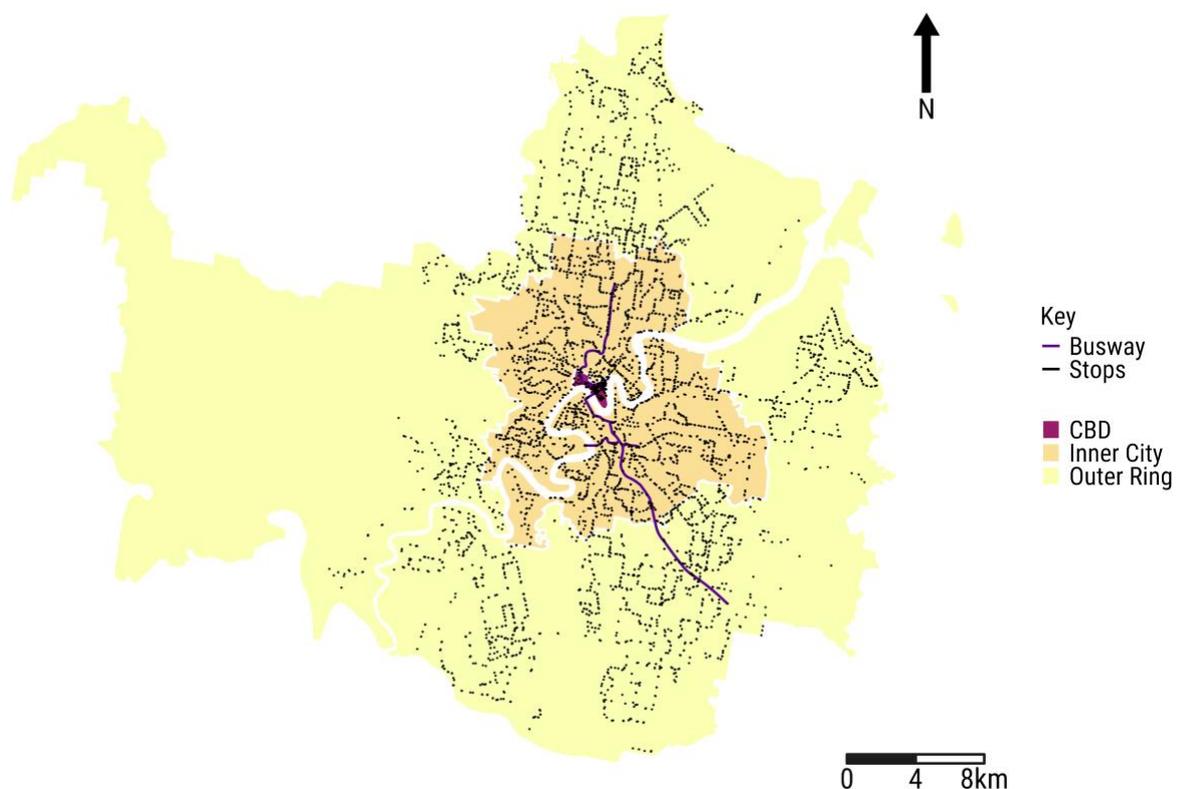

**Fig. 1.** Study Context

The weather in Brisbane can be categorised as subtropical, with a monthly day-time temperature over 23°C most time of the year. In summer (between December and February), the highest temperature can reach over 30°C during the day, while in winter (between June and August), the highest day temperature is around 20°C (Bureau of Meterology, 2021). Rainfall is most intense in the summer (monthly average rainfall over 130 mm), and less so during other periods of the year (ranging from 20 to 100 mm).

## 3.2. Data Sources

Timeframe, geographical coverage and information included in the data

### 3.2.1. Smart card data

For the current study, the main data source is a three-month smart card dataset of over 26 million records that stores trip-making records across South East Queensland by bus from March 1st to May 31st in 2016. For the purpose of the study, we focused only on trips within the Brisbane city council area. A trip record is generated every time a passenger taps on or off the bus at the start and end of a trip, respectively. The main information in a trip record includes boarding and alighting time (date, hour and minutes), stop name, service route, smart card ID and journey ID for linked trips. We geo-coded bus-trip records with the General Transit Feed Specification (or GTFS) and examined the trip durations. We removed trip records that could not be geo-coded due to missing or erroneous information of bus stops and those with overly long durations (i.e., over 3 hours), which represented rare and possibly, wrongly coded cases. The final dataset contained 10,661,040 hourly observations across 5226 stops within Brisbane.

### 3.2.2. Weather data

The weather data were drawn from the weather station of the University of Queensland, which was the only dataset available at the time of conducting this study. The weather data contained tracks of weather conditions (i.e., outdoor temperature, wind speed, relative humidity and rainfall) at a five-minute interval. For the current study, we aggregated the weather data and estimated average values (and sum for rainfall) of the weather conditions at the hourly level. We also estimated apparent temperature as a composite score of temperature, wind speed and relative humidity following the approach provided by the Bureau of Meteorology (2010) in Australia to approximate the level of thermal conform experienced by people. Figure 2 displays the temporal patterns of the weather variables over the study period.

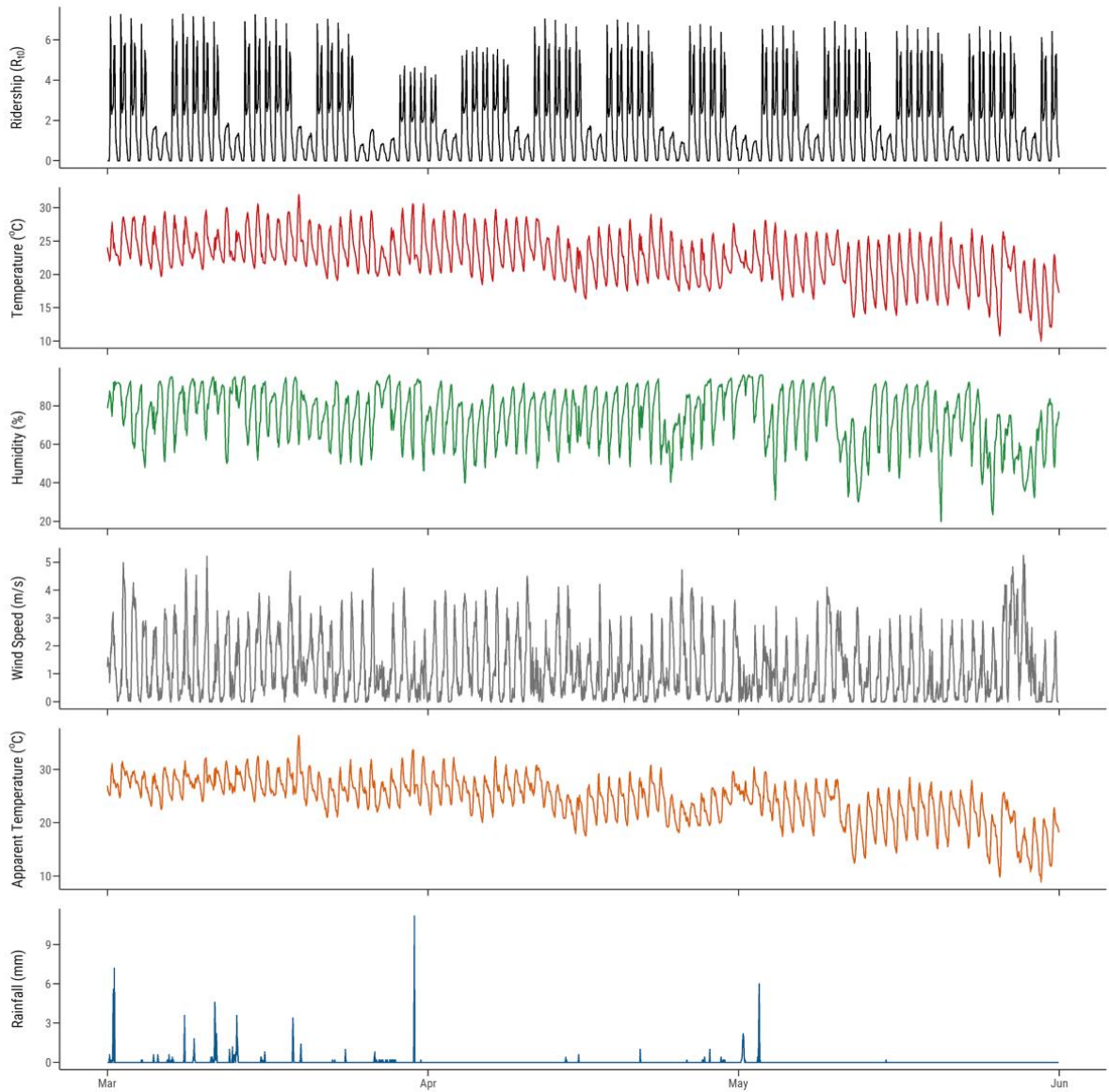

**Fig. 2**. Hourly weather variables, March 1st to May 31st, 2016.

3.2.3. OpenStreetMap data

We also drew on information from OpenStreetMap (OSM) to capture the influence of urban features on bus ridership. High density of key urban features, such as retail and educational amenities have been associated with high intensity of local travel activity (Graells-Garrido *et al.*, 2021). OSM is a crowdsourced database underpinned by a collaborative project to create a free editable map of the world. Users contribute geographic information. The dataset offers greater and more up-to-date coverage and accuracy than highly used commercial maps, such as datasets produced by Great Britain's national mapping agency (Haklay, 2010).

We extracted data on relevant urban features from OSM via API using the R package 'osmdata' (Padgham *et al.*, 2017). We identified 15 urban features (see details in Supplementary Material (SM) Table 1). Specifically, we measured the density of urban features based on a 400-metre buffer around each of the 5,226 bus stops in the transport network in Brisbane. This threshold was used because service planning guidelines for Australian cities specify that 90 percent of the local households need to be within the 400 metre walking distance from a bus stop (Daniels and Mulley, 2013). Additionally, literature suggests that 400 meters is a common rule of thumb for walking distance to access public transit stops in urban planning in Canadian and American cities (Foda and Osman, 2010; Hess, 2012).

3.2.4. Calendar event, service hours and stop location data

We also use two key sources to extract information on calendar events, bus services and stop locations. First, calendar events (e.g., weekdays, weekends, public and school holidays) were extracted from the Queensland's state calendar (DETE, 2013). Second, GTFS data (an open-source data containing service attributes of local transit services) were used to extract the service hours of bus (Google Developers, 2020) and geographic information of Brisbane's bus network, including the longitude and latitude of the bus stops. Such information was added to the smart card data based on the matching bus stop IDs in the GTFS, while weather variables were added based on the trip time.

## 4. Methodology

We developed a general model specification of hourly bus ridership at the stop level, to generate one hour-ahead predictions of bus ridership. Next we describe this specification before presenting the three machine learning approaches tested and metrics used to assess their bus ridership prediction outputs. Fig.3 illustrates the key steps of our methodology.

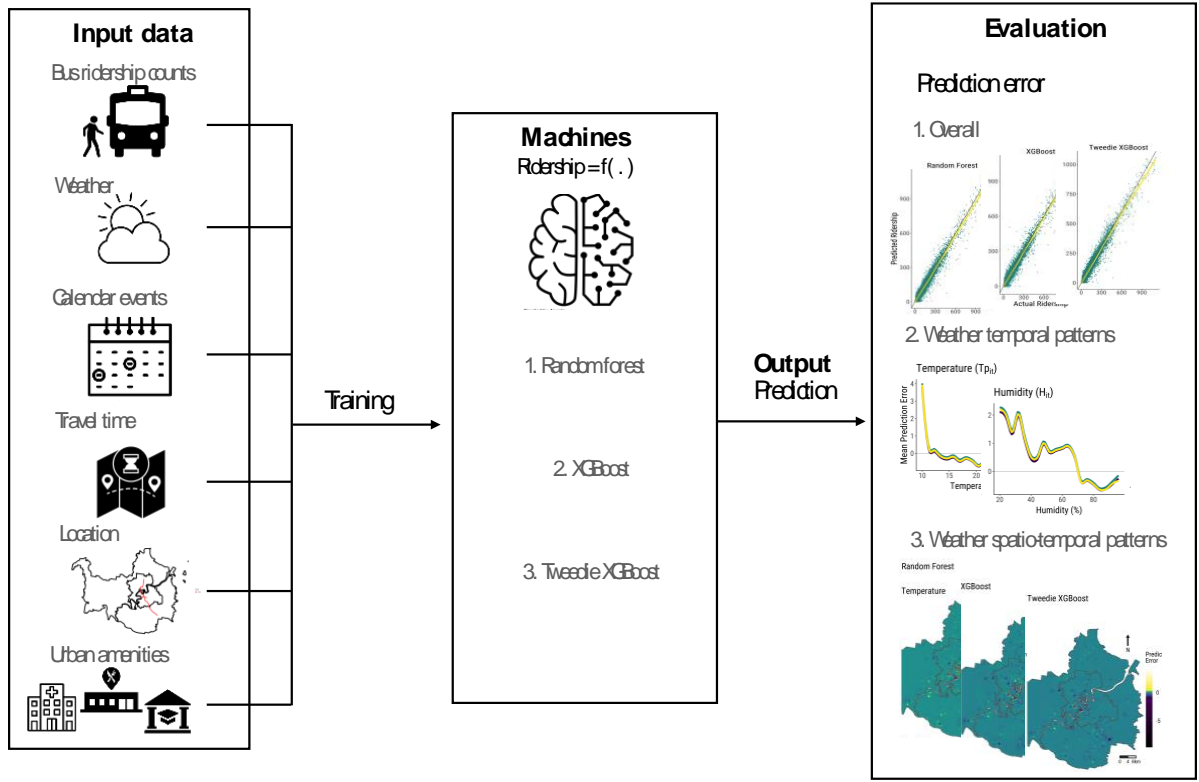

**Fig. 3**. Methodological framework.

4.1. General model specification

The model specification aims to generate prediction of bus ridership at hour *t+1* based on the ridership of the previous hour *t* at the stop level. A total of 43 independent variables were included in the modelling analysis, which can be written as:

$$R_{it+1} = f(R_{it} + R_{it-n} + W_{it} + C_t + L_i + G_i + F_i) + \varepsilon_{it} \quad (1)$$
where:
$$R_{it-n} = R_{it-24} + R_{it-168} \quad (2)$$
$$W_{it} = Tp_{it} + H_{it} + WS_{it} + AT_{it} + Rf_{it} \quad (3)$$
$$C_t = Hr_t + Dw_t + We_t + P_t + S_t + Fd_t + AM_t + PM_t + Wp_t + N_t + SI_t \quad (4)$$
$$L_i = Q1_i + Q2_i + Q3_i + Q4_i + Q5_i \quad (5)$$
$$G_i = CC_i + IC_i + IR_i + BW_i \quad (6)$$
$$F_i = Sn_i + Ed_i + U_i + Tr_i + Fn_i + Hc_i + SE_i + LE_i + NE_i + Rl_i + Cv_i + In_i + LF_i + Sh_i + Tr_i \quad (7)$$

where: $R_{it+1}$ represents bus ridership counts at bus stop *i* at time *t+1*; $f$ is a function of concurrent ridership $R$ and past ridership $R_{t-x}$, weather conditions $W$, calendar events $C$, trip duration $L$, geographical location $G$ and urban features $F$ at each bus stop; and $\varepsilon$ is a random error. Table 1 provides details of the full set of 43 independent variables included in our models. *Past bus ridership* (Eq.2) comprises a vector of two variables. Based on empirical evidence on the relationship between weather and bus ridership in Brisbane (Tao *et al.*, 2018), we included bus ridership lags at 24 and 168 hrs to account for temporal dependence and seasonal recurrent patterns of travel behaviour.

We assessed the impact of *concurrent weather conditions* (Eq.3) on bus ridership, capturing variations in temperature $Tp$, humidity $H$, wind speed $WS$, apparent temperature $AT$, rainfall $Rf$ and apparent temperature *AT*.

*Calendar events* were captured by a series of dummy variables including different times of the week (e.g., weekday and weekends), workdays and holidays. In addition, dummy variables were also created to measure differences in ridership over peak hours in the morning ($AM$), peak hours in the afternoon ($PM$) during working days, and during weekends ($Wp$). We also included a dummy variable to identify night-time hours ($N$) when travel demand is reduced, and when services were running ($SI$) to distinguish between *real* zero ridership counts and *false* zero ridership counts due to unavailability of bus service during certain time periods (e.g., midnight).

The *duration of travel* may also influence bus travel demand as a most convenient modal option (Phanikumar and Maitra, 2007; Tao *et al.*, 2017). Additionally, bus stops close to key activity destinations, such as the CBD, universities and shopping malls tend to attract trips from more distant areas (Tao *et al.*, 2014). We included the proportion of riderships according to different trip durations (Table 1). We identified irregular structural breaks in the distribution of journey time duration to define these proportions using Jenks partitioning (Jenks, 1977).

Systematic differences in bus ridership demand also exist across specific *areas of the urban space*. We added dummy variables to account for differences in bus ridership across city centre ($CC$), inner city ($IC$) and outer ring ($IR$) areas. The intensity of bus usage tends to diminish as we move away from cities' core employment centres (Tao *et al.*, 2014). We also included a dummy variable to identify bus stops along Brisbane's busway ($BW$). As the busway provides an exclusive highway for buses and sheltered bus stops, passenger demand along the busway corridor tends to be higher (Tao *et al.*, 2014).

Based on the OSM data described in Section 3.2.3, we included variables to measure variations in bus usage according to their accessibility to urban amenities. Higher accessibility to urban amenities has been associated with patterns of more localised and longer trips (Graells-Garrido *et al.*, 2021). Specifically we captured variability in walking distance accessibility to a range of 15 urban amenities across bus stops.

**Table 1.** Model variable description and descriptive statistics.

| | Description | Mean | Min. | Max. | Median | St. Dev. |
|---|---|---|---|---|---|---|
| *Ridership* | | | | | | |
| Ridership ($R_{it+1}$) | Bus ridership counts from stop i at time t+1hr | 1.64 | 0 | 1250 | 0 | 14.24 |
| Hour Lag ($R_{it}$) | Bus ridership counts from stop i at time t, temporal lag | 1.64 | 0 | 1250 | 0 | 14.24 |
| Day Lag ($R_{it-24}$) | Bus ridership counts from stop i at time t-24hrs, seasonal lag | 1.64 | 0 | 1250 | 0 | 14.28 |
| Week Lag ($R_{it-168}$) | Bus ridership counts from stop i at time t-128hrs, seasonal lab | 1.66 | 0 | 1250 | 0 | 14.4 |
| *Weather* | | | | | | |
| Temperature ($Tp_{it}$) | Temperature in degree Celsius | 22.4 | 9.99 | 31.97 | 22.5 | 3.6 |
| Humidity ($H_{it}$) | Ratio of absolute humidity to the highest possible absolute humidity, percentage | 73.84 | 20 | 96 | 75.87 | 14.56 |
| Wind Speed ($Ws_{it}$) | Wind speed in meter per second | 1.17 | 0 | 5.25 | 0.81 | 1.13 |
| Apparent Temp ($AT_{it}$) | The Steadman apparent temperature as defined by the Australian Bureau of Meteorology. | 24.17 | 8.89 | 36.35 | 24.84 | 4.21 |
| Rainfall ($Rf_{it}$) | Total rainfall depth in millimeters | 0.04 | 0 | 11.2 | 0 | 0.39 |
| *Calendar Events* | | | | | | |
| Hour of Day ($Hr_t$) | Identifier for the hour of the day | 12.5 | 1 | 24 | 12.5 | 6.92 |
| Day of Week ($Dw_t$) | Identifier for the day of the week | 3.98 | 1 | 7 | 4 | 2 |
| Weekend ($We_t$) | 1 if a bus ridership occurred in a weekend day | 1.28 | 1 | 2 | 1 | 0.45 |
| Public Holiday ($P_t$) | 1 if a bus ridership occurred in a public holiday | 1.06 | 1 | 2 | 1 | 0.24 |
| School Holiday ($S_t$) | 1 if a bus ridership occurred in a school holiday | 1.16 | 1 | 2 | 1 | 0.37 |
| Flexible Del Day ($Fd_t$) | 1 if a bus ridership occurred in a flexible staff professional development day | 1.04 | 1 | 2 | 1 | 0.18 |
| AM Peak ($AM_t$) | 1 if a bus ridership occurred during morning peak hours in a work day, 7-8am, Mon-Fri | 1.08 | 1 | 2 | 1 | 0.28 |
| PM Peak ($PM_t$) | 1 if a bus ridership occurred during afternoon peak hours in a work day, 15-17, Mon-Fri | 1.12 | 1 | 2 | 1 | 0.33 |
| Weekend Peak ($Wp_t$) | 1 if a bus ridership occurred during peak hours in weekends, 09-17, Sat-Sun | 1.38 | 1 | 2 | 1 | 0.48 |
| Night Hours ($N_t$) | 1 if a bus ridership occurred during night hours, 22-5am | 1.33 | 1 | 2 | 1 | 0.47 |
| Service Included ($SI_t$) | 1 if bus services are running | 1.39 | 1 | 2 | 1 | 0.49 |
| *Travel Time*[1] | | | | | | |
| JT Quantile 1 ($Q1_i$) | Share of bus riderships from stop i < 6.85 mins. | 0.26 | 0 | 1 | 0.15 | 0.28 |
| JT Quantile 2 ($Q2_i$) | Share of bus riderships from stop i between 6.85 and 11.68 mins. | 0.2 | 0 | 1 | 0.14 | 0.21 |
| JT Quantile 3 ($Q3_i$) | Share of bus riderships from stop i between 11.68 and 17.43 mins. | 0.17 | 0 | 1 | 0.12 | 0.18 |
| JT Quantile 4 ($Q4_i$) | Share of bus riderships from stop i between 17.43 and 25.5 mins. | 0.16 | 0 | 1 | 0.11 | 0.17 |
| JT Quantile 5 ($Q5_i$) | Share of bus riderships from stop i > 25.5 mins | 0.21 | 0 | 1 | 0.1 | 0.26 |
| *Location* | | | | | | |
| City Centre ($CC_i$) | 1 if boarding in the city centre | 1.03 | 1 | 2 | 1 | 0.17 |
| Inner City ($IC_i$) | 1 if boarding in the inner city | 1.49 | 1 | 2 | 1 | 0.5 |
| Outer Ring ($IR_i$) | 1 if boarding in the outer ring | 1.48 | 1 | 2 | 1 | 0.5 |
| Busway ($BW_i$) | 1 if boarding in the busway | 2.02 | 2 | 3 | 2 | 0.12 |
| *Urban Amenities*[2] | | | | | | |
| Sustenance ($Sn_i$) | Number of food and drink outlets within a 400m buffer from a bus stop | 0.22 | 0 | 1 | 0.19 | 0.23 |
| Education ($Ed_i$) | Number of education amenities (exc. univerities) within a 400m buffer from a bus stop | 0.2 | 0 | 1 | 0.07 | 0.28 |
| University ($U_i$) | Number of universities within a 400m buffer from a bus stop | 0.01 | 0 | 1 | 0 | 0.06 |
| Transport ($Tr_i$) | Number of transport amenities within a 400m buffer from a bus stop | 0.01 | 0 | 1 | 0 | 0.07 |
| Financial Services ($Fn_i$) | Number of financial services amenities within a 400m buffer from a bus stop | 0.01 | 0 | 0.5 | 0 | 0.03 |
| Healthcare ($Hc_i$) | Number of healthcare amenities within a 400m buffer from a bus stop | 0.07 | 0 | 1 | 0 | 0.15 |
| Small Day Ent ($SE_i$) | Number of small day entertainment venues within a 400m buffer from a bus stop | 0.02 | 0 | 1 | 0 | 0.07 |
| Large Day Ent ($LE_i$) | Number of large day entertainment venues within a 400m buffer from a bus stop | 0 | 0 | 0.5 | 0 | 0.02 |
| Night Ent ($NE_i$) | Number of night ent amenities within a 400m buffer from a bus stop | 0 | 0 | 0.33 | 0 | 0.01 |
| Religious ($Rl_i$) | Number of religious amenities within a 400m buffer from a bus stop | 0.08 | 0 | 1 | 0 | 0.17 |
| Civic ($Cv_i$) | Number of civic amenities within a 400m buffer from a bus stop | 0.01 | 0 | 1 | 0 | 0.06 |
| Infrastructure ($In_i$) | Number of infrastructure amenities within a 400m buffer from a bus stop | 0.02 | 0 | 1 | 0 | 0.07 |
| Leisure ($L_i$) | Number of leisure amenities (eg. sauna, GYMs) within a 400m buffer from a bus stop | 0.1 | 0 | 1 | 0 | 0.2 |
| Shops ($Sh_i$) | Number of shops amenities within a 400m buffer from a bus stop | 0.21 | 0 | 1 | 0.14 | 0.24 |
| Tourism ($Tr_i$) | Number of tourism amenities within a 400m buffer from a bus stop | 0.04 | 0 | 1 | 0 | 0.14 |

[1] Travel time ranges: Jenks partitioning was used to identify natural breaks structural time breaks in the distribution of journey time duration across the bus network.
[2] Urban amenities: We measured a 400m buffer based on service planning guidelines for Australian cities (Daniels and Mulley 2013).

## 4.2. Machine learning algorithms

We assessed three tree-based machine learning algorithms - RF, XGBoost and Tweedie XGBoost algorithms. We compared the overall predictive model performance and predictive model performance according to changes in weather conditions. Tree-based machine learning algorithms were employed as they provide an appropriate balance between interpretability and model complexity (James *et al.*, 2013). Their flexible non-parametric structure offers the potential to identify functional spatial and temporal structures, and novel approaches have been developed to visualise and interpret their outputs (Molnar, 2021). Moreover, tree-based machine learning algorithms have successfully been used to predict overly dispersed data count in a variety of fields (e.g. Yan *et al.*, 2010; Weng *et al.*, 2018); that is, data with similar attributes to those often used in urban transport research. Tree-based machine learning techniques are thus preferred over more flexible but complex machine learning approaches, such as deep neural networks. Developing a reliable neural network model requires careful specification of appropriate hidden layers, weights between input and output layers, activation, optimisation forward and back propagation functions (Skapura, 1996). While they might provide better predictions, they are: computationally expensive; complex to train, customise and deploy; and, hard to interpret (James *et al.*, 2013). The remainder of the section offers a brief description overview of each of the approaches we use to estimate Eq (1), and their relative advantages and disadvantages.

### 4.2.1 Random Forest (RF)

RF is one of the most widely applied machine learning techniques due to its ease of use and overall performance (Breiman, 2001). It is a combination of tree models in which each tree depends on the values of a random vector sampled independently, which are de-correlated. RF uses bootstrap aggregation to combine outputs from a series of models into a single final prediction, with the final model prediction being the average across all individual trees. RF has key advantages over other three-based machine learning algorithms. It generally has a high prediction accuracy, is robust to overfitting, can handle outliers and noise, and scale efficiently to large datasets (e.g. Krauss *et al.*, 2017; Ahmad *et al.*, 2018; Misra and Wu, 2020).

### 4.2.2 eXtreme Gradient Boosting (XGBoost)

XGBoost is also an ensemble algorithm that combines outputs from multiple models to produce a single prediction, and represents an adaptation of the gradient boosting machine algorithm proposed by Friedman (2001). As a form of gradient boosting, XGBoost utilises gradient descent to improve model performance, and decision trees are built iteratively, with each tree built to minimise the error residuals of its predecessor. Yet, it has been optimised for scalability and computational efficiency, allowing it to achieve high predictive accuracy with minimal training time (Chen *et al.*, 2016). It has become widely recognised as one of the most effective machine learning models available and is a prominent off-the-shelf data mining technique in machine learning competitions (e.g. Chen *et al.*, 2016).

### 4.2.3 Tweedie XGBoost

We also used a novel XGBoost version, known as Tweedie XGBoost, to address two key challenges: overdispersion and zero inflation in the data. These challenges are generally present in transport research, but they are rarely addressed explicitly in machine learning applications in predicting urban passenger counts. Yet, they can significantly reduce prediction accuracy at locations with high ridership demand as the algorithm is able learn to more accurately predict events with high frequency occurrence (i.e. zero or small value counts) in the data, and produces less accurate predictions at locations of high demand. However, the latter may also need more precise predictions.

Tweedie XGBoost extends the XGBoost algorithm (which is based on a Gaussian distribution) to incorporate Tweedie distributions. These distributions offer a family of general functional forms, known as exponential dispersion models. Tweedie distributions include a class of compound Poisson–gamma distributions which enable effective modelling and prediction of non-negative data with a high density of zero counts (Gilchrist and Drinkwater, 2000). A Poisson–gamma distribution can be adjusted to better fit the data by specifying two parameters: its variance power (VP) and link power (LP) functions. To find the most optimal approximation for our data, we assessed a range of VP between 1 and 2 and set the link power as the canonical link function of 1 - VP. We then ran a series of models using a range of VPs between 1 and 2, and used the best performing model to generate bus ridership predictions.

### 4.3. Implementation

We used the open-source machine learning and artificial intelligence platform '*H2O.ai*', to fit and assess our models in the R environment. H2O.ai enables handling large data sets, efficient parallelisation and provides a coherent framework to build a variety of machine learning approaches to produce consistent model outputs. For the modelling analysis, we randomly split the data into a training set (80%; over 8.5m) and a validation set (20%; over 2.1m); used 5-fold cross validation to train models; and, assessed their prediction accuracy based on a suite of prediction error measures. We based these measures on the unseen validation dataset. Unseen test data are used for validation because we are interested in the ability of a model to generate accurate predictions on datasets which they model has not seen. We used a random grid search to tune the set of optimal hyper-parameters for our models. For each algorithm, the best performing model in terms of prediction error based on the mean square error was selected for a final cross-model assessment comparison. Table 2 reports the hyperparameters used for our final set of models. All our models were fitted in a Dell Precision Tower with 80 RAM and 28 cores.

**Table 2**. Hyper-parameters used for our final set of models.

| Parameters | Random Forest | XGBoost | Tweedie XGBoost |
| --- | --- | --- | --- |
| Training sample size | 0.8 | 0.8 | 0.8 |
| Learning rate | – | 0.3 | 0.05 |
| Number of trees | 76 | 55 | 1000 |
| Number of variables | 13 | – | – |
| Maximum depth | 30 | 3 | 10 |
| Minimum observations | 3 | 5 | 1 |
| Number of bins | 25 | – | – |
| Row sample rate per tree | 0.95 | 0.55 | 0.55 |
| Column sample rate per tree | – | 0.3 | 0.9 |
| Gamma | – | 1e-8 | 1e-4 |
| Tweedie Power | – | – | 1.06 |

*Note:*
Learning rate: Shrinkage parameter at each iteration determining the step size in the loss function.
Number of trees: Number of trees to build.
Number of variables: Number of randomly sampled variables.
Maximum depth: Maximum tree depth.
Minimum observations: Minimum number of observations for a leaf.
Number of bins: Number of bins for the histogram to build, then split at the best point.
Row sample rate per tree: Number of row sampling rate.
Column sample rate per tree: Number of column sampling rate.
Gamma: Minimum relative improvement in squared error reduction for a split.
Tweedie Power: Parameter to define the distribution function needed for a Tweedie regression.

## 5. Results

In this section, we first discuss the overall predictive performance of our models before analysing the temporal and spatio-temporal patterns of the prediction error. We then focus on assessing the predictive capacity of our models during hours of extreme weather readings.

5.1 Overall prediction performance

Table 3 reports error indicators to assess the overall performance of our models. Fig.4 displays the association between observed and predicted bus ridership. Collectively, the reported metrics point to similarly high levels of overall prediction accuracy for our three models: low RMSE, a narrow interquartile range, high correlation between observed and predicted ridership, over 81% of zero prediction error, and over 99% of prediction errors within 1.5 standard deviations around the mean error (i.e. errors within -/+22 passengers) - which is statistically acceptable. However, larger differences in computational time exist across our three models. The training time for the Tweedie XGBoost is 2.7 times that for the standard XGBoost and 3.5 times that for the RF. These preliminary results favour the RF providing high prediction accuracy and requiring less computing time.

**Table 3.** Comparison of Model Performance

| Performance Indicator | Random Forest | XGBoost | Tweedie XGBoost |
|---|---|---|---|
| Test RMSE | 2.497 | 2.488 | 2.447 |
| Interquartile range | -0.144 | -0.117 | -0.093 |
| *Test prediction error* % | | | |
| -100+ | 0.004 | 0.004 | 0.004 |
| -100 – -51 | 0.018 | 0.016 | 0.017 |
| -50 – -23 | 0.082 | 0.084 | 0.082 |
| -22 – -1 | 9.238 | 9.248 | 9.165 |
| 0 | 81.309 | 81.605 | 81.983 |
| 1 – 22 | 9.283 | 8.955 | 8.673 |
| 23 – 50 | 0.057 | 0.075 | 0.063 |
| 51 – 100 | 0.009 | 0.012 | 0.012 |
| 100+ | 0.001 | 0.001 | 0.001 |
| Pearson correlation | 0.985 | 0.985 | 0.985 |
| Training time (mins) | 50.522 | 65.599 | 176.462 |

*Note:*
Test RMSE: Root Mean Squared Error based on test sample.
Interquartile range: Quartiles of model prediction error based on test sample.
Pearson correlation: Correlation between observed and predicted bus ridership.

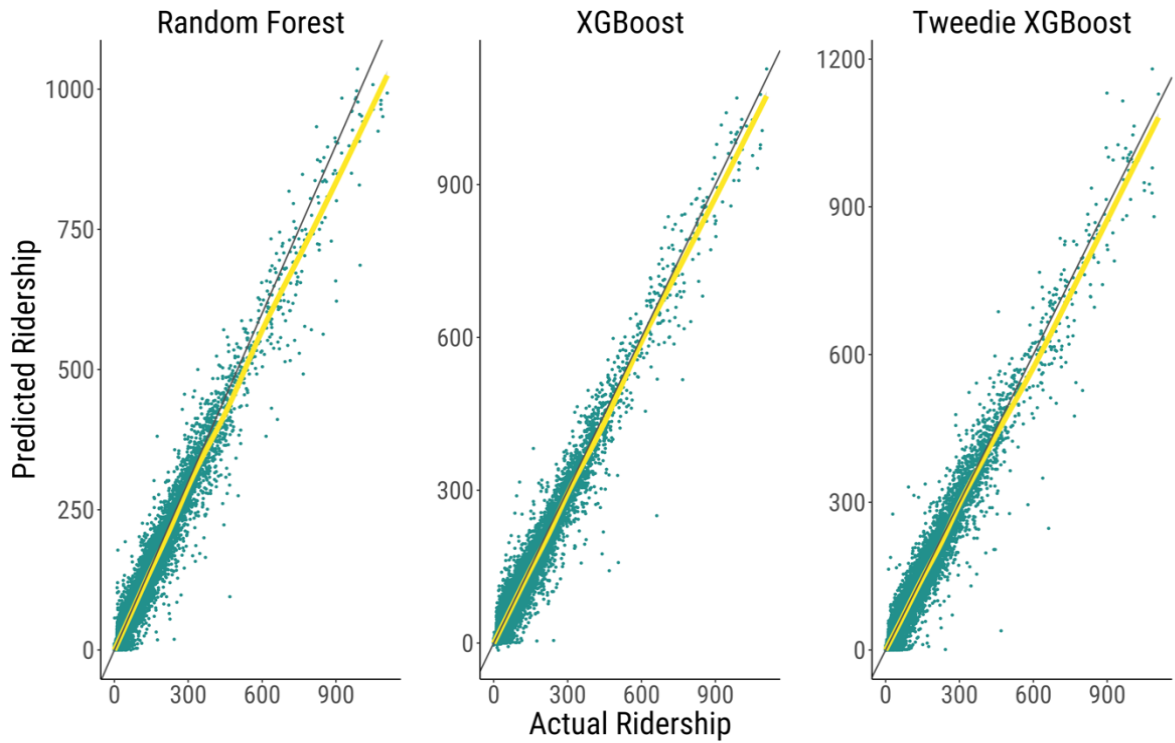

**Fig. 4.** Scatter plots of observed and predicted ridership ($R_{t0}$) based on unseen test data. The black line indicates perfect correspondence. The yellow line represents a local polynomial regression estimated by locally weighted scatterplot smoothing (loess) to capture the correspondence between observed and predicted ridership.

Generally, weather variables are less influential than other factors, such as certain seasonal ridership dependencies, urban amenities and calendar events in predicting (see SM Fig. 1). However, it is likely weather may induce considerable spatio-temporal variability in bus ridership demand across the Brisbane bus network, particularly in activity-intense areas, including the CBD, university and shopping centres (Tao *et al.*, 2018). Additionally, we know that extreme weather events are rare occurrences, and hence their influence in overall prediction tends to be small. The influence of weather is likely to be highly localised in space and time. Examining the spatial and temporal variations of impacts and prediction of bus ridership in response to weather conditions is thus to key to the local operation of bus service.

5.2 Temporal patterns of bus ridership prediction

We examined the temporal patterns of prediction accuracy by measuring the mean hourly prediction error (Fig.5). The results show a generally small range of mean hourly prediction error for all three models, although some differences exist. The Tweedie XGBoost and XGBoost display a consistently narrow prediction error range, while large fluctuations are observed for the RF, particularly on May 9th with prediction error exceeding -2 during a period of a pronounced drop in apparent temperature (see Fig.2). Similarly notable prediction errors are observed on April 4th, 11th and 26th, all representing underpredictions during peak hours of weekdays. These patterns point to slightly more robust predictions from the XGBoost and Tweedie XGBoost algorithms, especially in a context of sudden weather fluctuations.

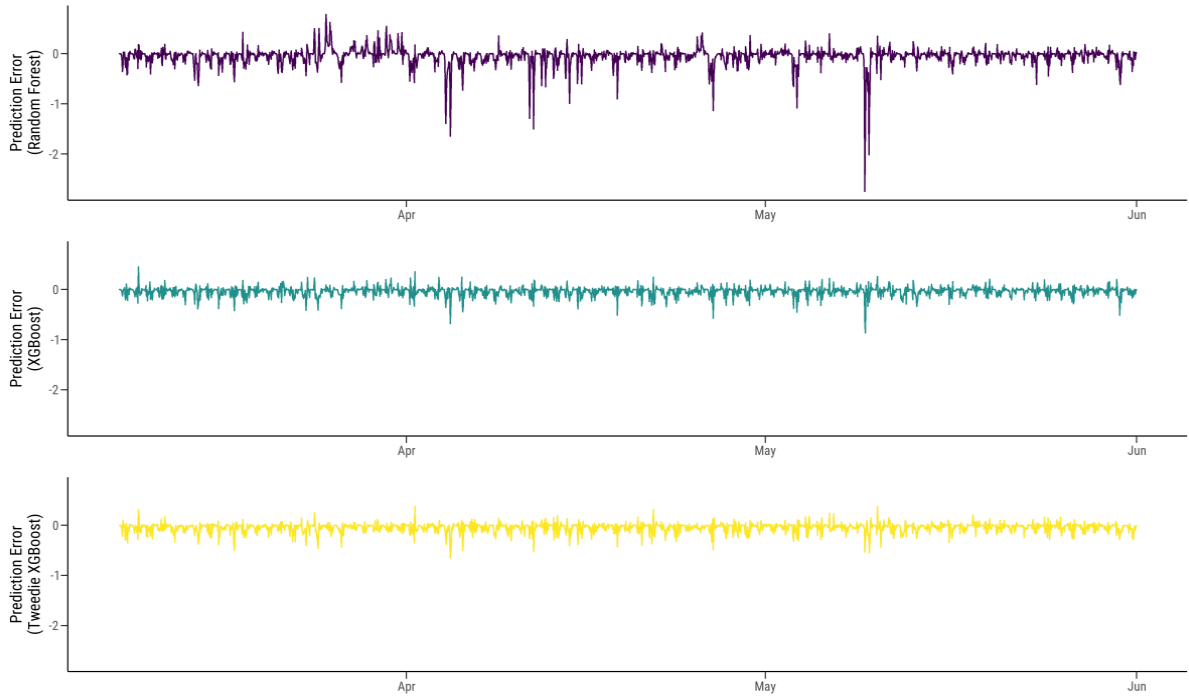

**Fig. 5.** Mean hourly stop-level prediction error. Prediction error: predicted ridership minus observed ridership

5.3 Spatio-temporal patterns of bus ridership prediction

To assess spatio-temporal patterns of model accuracy, we averaged and mapped prediction errors (predicted minus observed ridership) by bus stop across the network and by relevant temporal slots, to examine errors for peak *versus* off-peak hours, and weekday *versus* weekend hours. Spatially continuous surfaces of these errors were generated and visualised using inverse distance weighted (IDW) interpolation via the R package 'gstat' (Pebesma and Graeler, 2021). Fig.6 displays these error surfaces with positive scores representing overprediction and negative scores encoding underprediction.

Examining predictions for all hours, Fig.6 reveals a small average stop-level error range moving from -4 to 1, and a systematic pattern of underprediction across much of the bus network. Prediction errors tend to be highly geographically localised. The largest overpredictions tend to occur in activity-intense areas, including the CBD, university and retail centres as well as artery roads and along the bus busway. Underprediction tends to occur in areas of lower activity intensity.

Across models, differences in all hours prediction errors exist. RF prediction errors display a wider value range, particularly underpredicting bus riderships, compared to XGBoost and Tweedie XGBoost prediction errors. Slight differences are observed in the geographic pattern of errors. The Tweedie XGBoost display a consistent pattern of higher accuracy in activity-intense locations where both the RF and XGBoost record overprediction, particularly in the CBD, South Bank, University of Queensland (UQ) Chancellor's Place stop and Chermside shopping mall in the north end of the busway.

Generally similar geographic patterns in prediction errors exist for peak *versus* off-peak hours. Underprediction is observed across most parts of the bus network, and overprediction in activity-intense areas. The extent of error is however consistently larger for peak hours predictions, particularly in the CBD and along certain commuter routes. Additionally, the error range is wider for the RF while narrower for the XGBoost, pointing to slightly greater overall accuracy for the XGBoost, particularly during peak hours.

For weekdays, prediction error is consistently larger than for weekends. However, the level of overprediction is higher for weekends than for weekdays, which is more pronounced around activity-intense areas relating to employment, education and retail centres. Again, the RF displays a larger error range particularly for weekdays. The Tweedie XGBoost exhibits a wider range for weekends, while large errors can be observed around certain retail centres, such as Indooroopilly and Garden City. The evidence thus far suggests a slight superiority of the standard XGBoost model over the Tweedie XGBoost and RF. The RF appears as the worst performing model. Yet, our primary goal is to achieve high accuracy predictions in the context of extreme weather conditions.

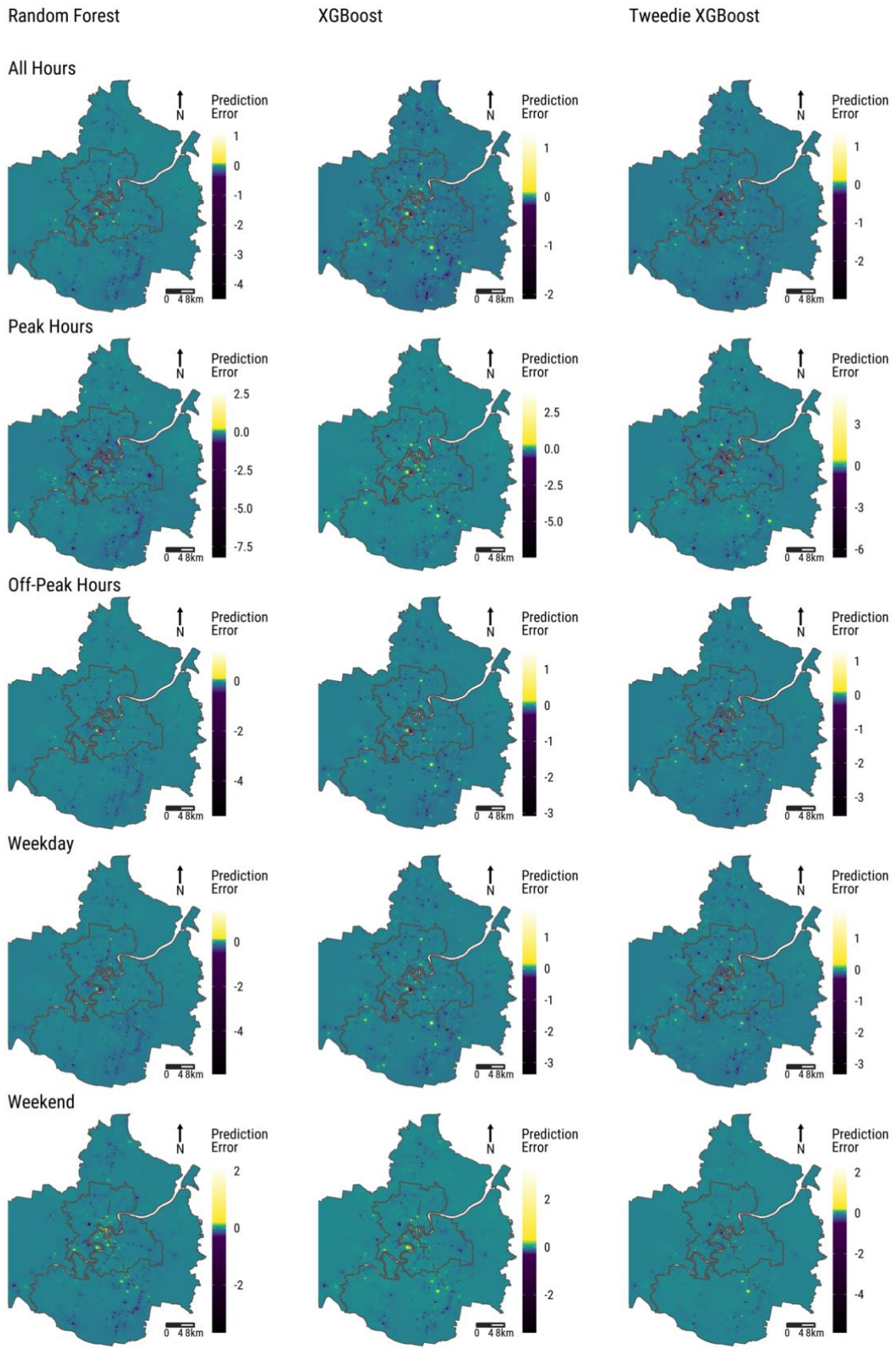

**Fig.6**. Average stop-level prediction error: predicted ridership minus observed ridership. Negative prediction error represents underprediction. Positive prediction error represents over prediction. Legend keys reflect the minimum and maximum values of the original data. Separate legends are used to better assess the extent of prediction error across models.

5.4 Bus ridership prediction during extreme weather

We examined the dependency of prediction error according to observed weather recordings for our five weather variables. Fig.7 shows how the prediction error changes with weather. Temperature and apparent temperature display relative high variability in prediction error. Errors are the largest for recording of less than 10°C and very small for higher temperatures (12-25°C). Similarly, prediction errors are higher for lower levels of humidity (<40%). Low temperature and humidity conditions are atypical in Brisbane where average day-time temperature revolves around 23°C and the average annual relative humidity is 66.8%. High rainfall is also associated with high prediction error, probably due to the rare and sudden nature of these events in subtropical climates. The associations between weather variables and prediction error are remarkably consistent across our three models.

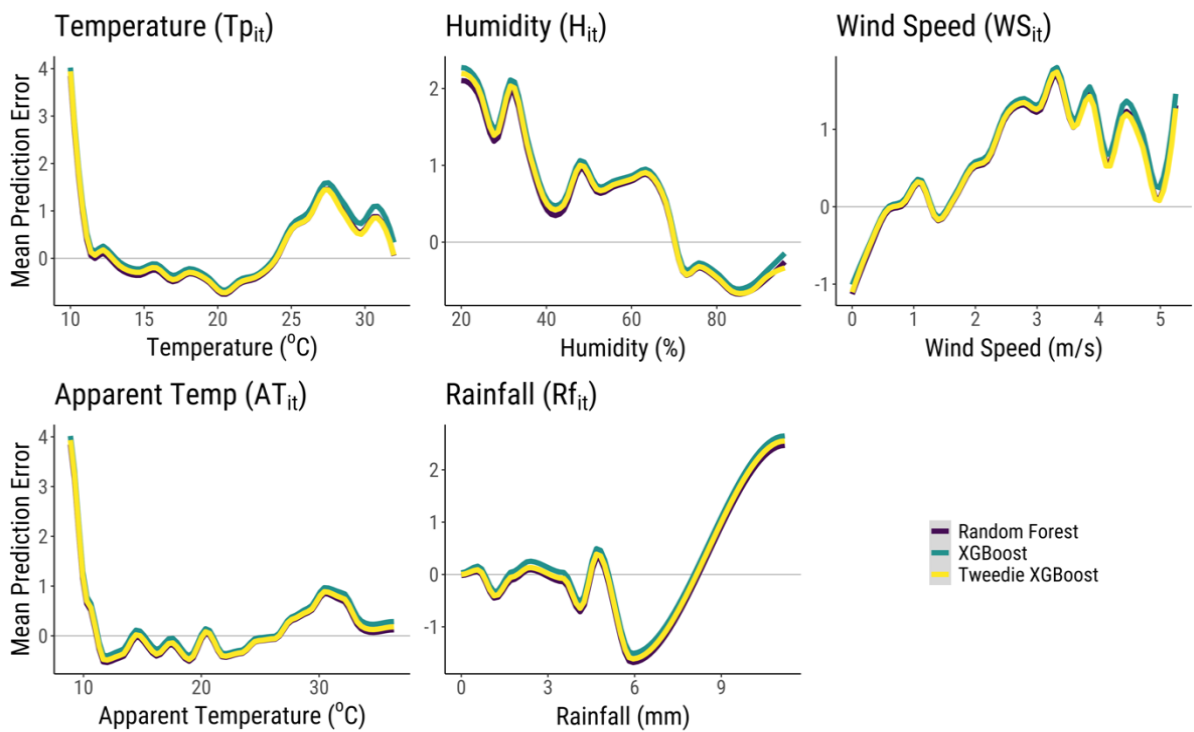

**Fig.7.** Mean prediction error for changing weather conditions. Prediction error: predicted ridership minus observed ridership

To analyse the spatial patterns of the weather-prediction error association, we examined prediction errors for observations during periods of "extreme" weather conditions. For each weather variable except rainfall, we defined any observations above or below 1.5 standard deviations of the average value as extreme weather readings. For rainfall, we considered any observations above 3mm as an extreme condition. Based on these sub-samples, we computed stop-specific prediction errors for each weather variable and created spatially continuous surfaces using IDW interpolation to display the results (Fig.8).

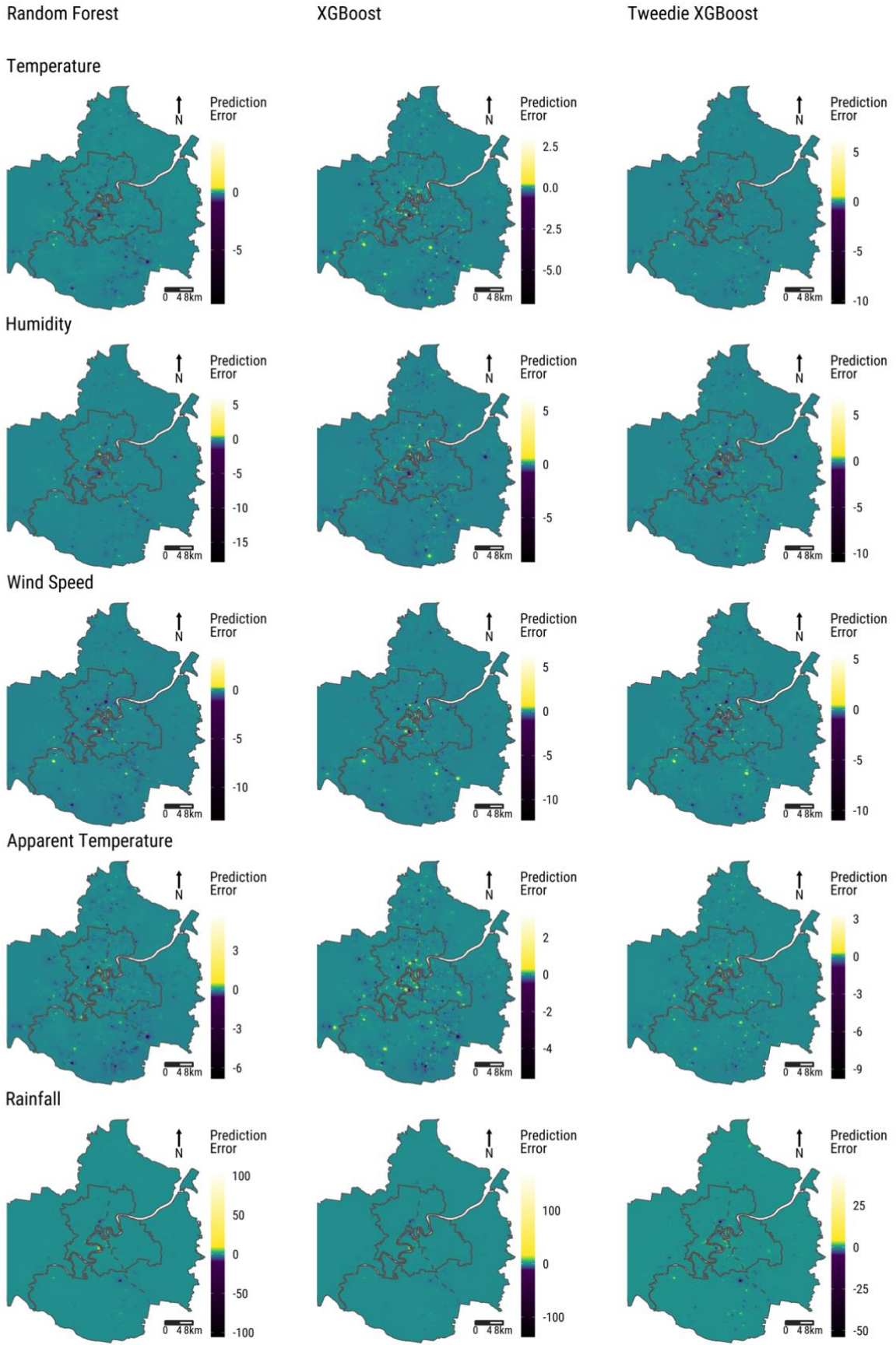

**Fig. 8**. Prediction error on unseen data during periods of extreme weather observations. Prediction error: predicted ridership minus observed ridership. Negative prediction error represents underprediction. Positive prediction error represents over prediction.

Fig.8 reveals that prediction accuracy varies across weather variables and models. The Tweedie XGBoost displays larger error ranges than the XGBoost for temperature, humidity and apparent temperature, and the RF for temperature and apparent temperature. These results point to worse overall prediction accuracy for the Tweedie XGBoost for temperature, humidity and apparent temperature. However, error ranges relating to these variables are relatively small ranging from around -15 to a maximum of 3.5. For rainfall, prediction errors display a much wider range, while the Tweedie XGBoost produced greater prediction accuracy, with errors ranging between around -50 to just over 25. For the RF and XGBoot, the errors vary from below -100 to over 100. Such gain in prediction accuracy arguably outweighs the smaller losses in accuracy by the Tweedie XGBoost predicting bus ridership, given also that predictions during extreme weather conditions is a major limitation across all models.

To better understand how extreme weather influences ridership and relates to the spatial variability in prediction observed in Fig.9, we analysed the observed patterns of ridership during "extreme" and "normal" weather conditions (i.e. all "non-extreme" weather observations). We used the difference in observed ridership between the periods of extreme and normal weather conditions. Positive differences indicate larger ridership counts during extreme than during normal weather, negative denoting the reverse.

Fig.9 displays the resulting differences. Systematic differences exist between observed ridership during extreme and normal weather conditions. The largest differences occur along certain commuter routes as well as around the CBD, university campuses and large retail centres. Periods of very low temperature, low apparent temperature and high humidity coincide with dramatically reduced ridership at stops located near retail and education centres, such as the CBD, university campuses and Sunnybank shopping district. Meanwhile, increased ridership is observed on certain commuter routes, such as busway stations and along artery roads. A similar yet reversed pattern is seen during periods of very high temperature, high apparent temperature, low humidity and high wind speeds. These periods coincide with higher ridership at stops near retail and education centres than an average "normal" day, whilst the ridership along commuter routes tend to remain similar to the all-hours average.

These patterns seem to reflect the discretionary nature of trips to retail and education centres, which tend to be more flexible and subject to change under uncommon conditions (e.g. more trips under pleasant weather and less so under unpleasant weather conditions). Meanwhile, mandatory trips, such as commuting, are less influenced by weather. However, less pleasant weather conditions may cause people who would normally walk or cycle to take the bus, resulting in increased ridership along commuter routes during these periods.

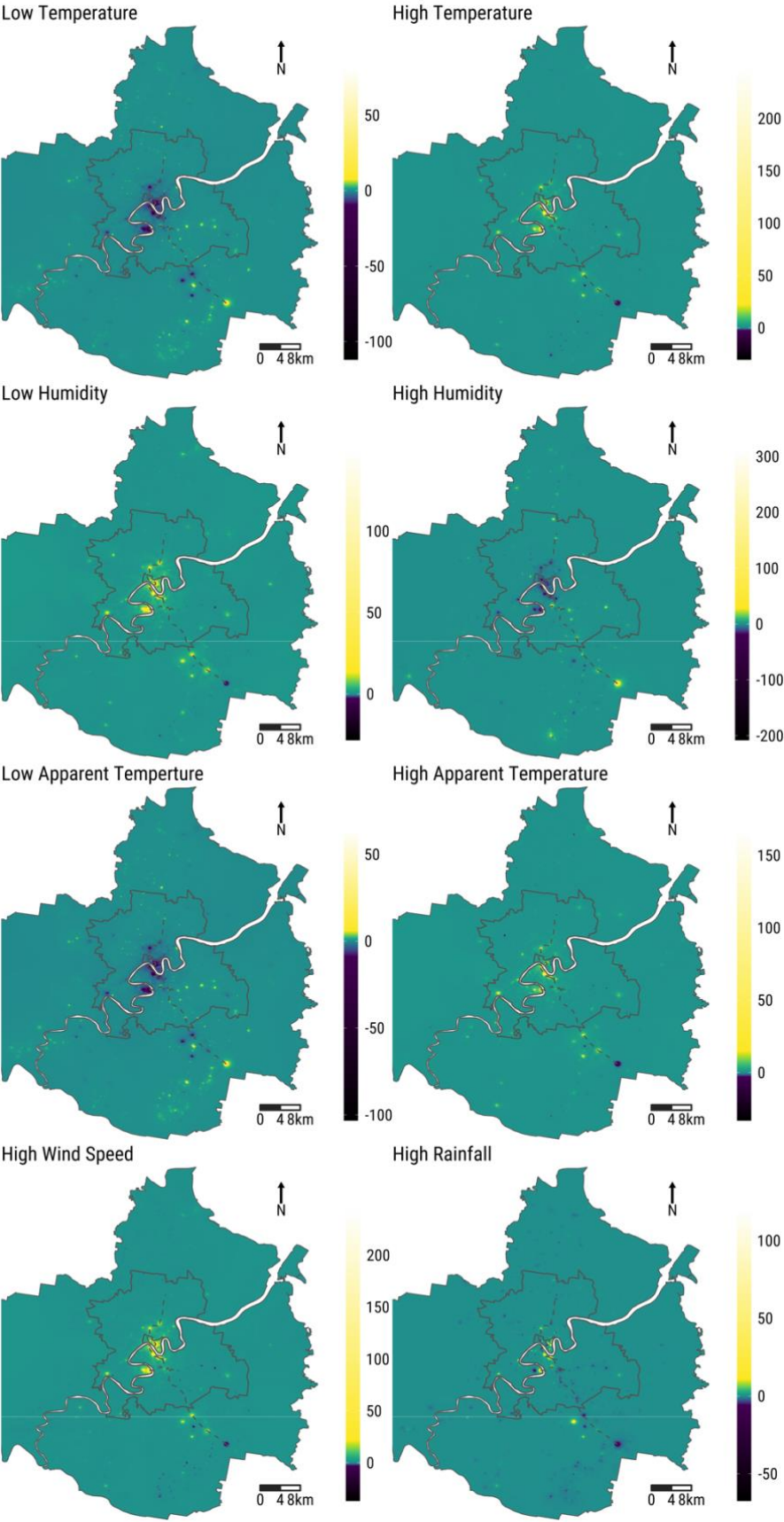

**Fig.9**. Difference between the average of observed bus ridership during periods of "extreme" and "normal" weather.

## 6. Discussion

6.1 Key results

This study provides a novel application of three machine learning algorithms - RF, XGBoost and Tweedie XGBoost - to assess their capacity to produce accurate one-hour ahead, stop-level bus ridership predictions during periods of sudden weather change. While we found evidence of relatively high accuracy for the three models, we also presented evidence of marked spatio-temporal variability in prediction error across models pointing to complex relationships between stop-level bus ridership and local weather conditions. Analyses of overall performance measures favoured RF predictions achieving a good balance between prediction accuracy and computing time. However, analyses of the spatial and temporal patterns of prediction error revealed the limitations of our RF in generating accurate stop-level bus ridership predictions in the context of extreme weather conditions. The RF also displayed relatively wide error ranges for predictions over periods of sudden fluctuations in apparent temperature. This reflects a more limited ability for the model to capture the combined impact of unusually low or high levels of extreme temperature and humidity during peak hours of weekdays.

Additionally, we presented evidence favouring the XGBoost and Tweedie XGBoost models to generate more accurate stop-level bus ridership predictions than the RF model. Yet, they also displayed strengths and weaknesses in different weather scenarios. The XGBoost showed consistently narrower error ranges during spells of extreme temperature, apparent temperature and humidity readings, but performing poorly during periods of heavy rainfall. By contrast, the Tweedie XGBoost demonstrated much greater accuracy predicting ridership during periods of heavy rainfall, but performed less well at generating predictions under the conditions of extremely high and low levels of temperature, apparent temperature and humidity. We also revealed patterns of systematic geographic variability in bus ridership during extreme weather conditions. Significant reductions in bus ridership were associated with retail and education centres during periods of adverse weather, while increases in ridership were observed along certain commuting routes.

6.2 Interpretation

Based on the analysis, the Tweedie XGBoost model emerged as the overall best-performing model. This algorithm displayed much narrower error ranges (from around -50 to 25) during periods of heavy rainfall compared to the XGBoost model (from below -100 to over 100), while differences in error ranges for other weather variables between the two models were largely negligible (i.e. less than 5 riderships). In Brisbane and the wider South East Queensland, rainfall is likely to continue to play a key role in influencing short-term bus ridership, as heavy rainfall is projected to become more frequent in the near future (Queensland Government, 2020). Rainfall can exert comparatively larger impacts than other weather variables on bus demand across the city's transit network, particularly in activity-intense locations such as the CBD, university and retail centres (Tao *et al.*, 2018). Our modelling and prediction results provide evidence base, to inform the operation of bus service and better cope with heavy rainfall and induced change in ridership.

We also presented evidence of the limitations of relying on overall measures of prediction accuracy for comparative model assessment. Normally overall measures of prediction performance in transport studies, such as the mean absolute percentage error (MAPE) and mean relative error (MRE) (Varghese *et al.*, 2020). Yet, these measures can conceal significant differences in prediction accuracy over space and time. We revealed significant variability in prediction errors across Brisbane's bus network and marked variations during weekdays and peak hours than weekends and off-peak hours. Based on space-time-specific error measures, XGBoost models were shown to outperform the RF in generating more accurate localised predictions of bus ridership.

Our findings also revealed marked geographic variability in prediction accuracy across the bus network associated with extreme weather conditions. Weather variables have highly localised and spatially heterogeneous effects on bus ridership demand, leading to higher or lower demand than expected - as reflected by large prediction error in certain stops. This is especially the case in activity-intense areas such as retail and education centres, where discretionary trips may be rescheduled, rerouted or cancelled. Commuting journeys may see increases due to modal shifts from walking or cycling to bus travel. This is consistent with previous research (Tao *et al.*, 2018) indicating that sudden changes of temperature and rainfall may exhibit markedly varying impacts across the bus network. The prominence of these factors on influencing local travel patterns are likely to increase due to climate change as annual average temperatures are expected to rise as well as the frequency of hot days, duration of warm spells, and intensity of heavy rain (Queensland Government, 2020).

Taken together, these findings suggest that real-time weather information should be incorporated in the monitoring of the bus transit demand, in addition to other conventionally relevant information such as built environment and calendar events. Although changes in weather may not always lead to adjustments to public transit service, real or near-real time bus demand predictions may better equip transit operators to make timely changes, especially in the face of sudden changes in weather. Services may be redeployed from bus stops experiencing decreases in ridership under adverse weather conditions to areas of increased service demand.

6.3. Challenges and limitations

Our analysis focused on a relatively short period of time (three-month) in a subtropical climatic context. Drawing on a longer time series could assist in developing algorithms with greater stop-level prediction accuracy. Machine-learning algorithms trained on a larger dataset covering seasonal variations in weather should be able to more precisely recognise observations of extreme weather, leading to more accurate predictions during these events. Such algorithms could also enhance our understanding of the varying impacts of weather on travel behaviour during different weather seasons across the transit network. The extent and geographic patterns of weather effects on travel behaviour may differ across cities (Koetse and Rietveld, 2009). Climatic zones differ. Human bodies and minds adjust and develop different expectations and tolerance to weather. Future research could improve our understanding of the real-time relationship between weather and travel behaviour by analysing this association in different geographical contexts. Such understanding is key to guiding context-specific planning strategies.

Our findings suggest that the behavioural response of bus passengers is complex and geographically highly localised. In addition to periodic travel patterns, urban amenities, calendar events and geographic location, socio-economic neighbourhood features and bus user profiles tend to influence bus ridership patterns. Explicitly incorporating all these factors into the modelling of weather influencing bus usage will likely provide a more thorough understanding of individuals' bus use decision making process under changing weather conditions, although privacy and confidentiality considerations in accessing information on individual user profiles will represent a challenge. We also used three off-the-shelf machine learning approaches of fast deployment. Recently sophisticated deep learning architectures of fast deployment have recently been designed to capture spatio-temporal dependences in traffic flow data (Zhou *et al.*, 2021). Future research could seek to adapt these models to predict bus ridership data.

Our analysis focused on a single mode of transport (bus). For future research, the use of different public transport modes, such as shared bicycle, ferry and train could be examined and integrated into predictive models. This, we expect, will provide more accurate insights into modal shifts under weather changes. Such analysis could greatly enhance our understanding of the dynamics of modal substitution during inclement weather, ultimately enhancing the capacity of transit operators to develop appropriate transport and planning responses.

## 7. Conclusion

Climate change has increased the frequency and severity shifts in extreme weather. These volatile weather conditions can have disruptive impacts on urban transport systems, affecting the demand for and usage of public transport for daily travel purposes. Drawing on a dataset of over 10 million observations, we assessed three machine learning algorithms - Random Forest (RF), eXtreme Gradient Boosting (XGBoost) and Tweedie XGBoost algorithms – to generate one-hour ahead predictions of bus ridership at the stop level during extreme weather events. Our results suggest that the Tweedie XGBoost generates the most accurate predictions, providing narrower prediction error intervals in a wide set of scenarios, particularly during periods of heavy rain. The proposed model has the capacity to account for the skewed distribution of bus ridership, and generate accurate real-time predictions. We contend that the current study represents a necessary step towards further channelling big-data analysis into the real-time operation of transit service in the face of an increasingly volatile climatic context (e.g., reallocating bus services). Future research may contribute to this area by comparing more advanced machine- and deep-learning algorithms in their capacity and efficiency of modelling weather-ridership interactions.

**Acknowledgments**

No funding was provided for this research. We would like to acknowledge the developers of the following R libraries used in our paper (order alphabethically): chron, doParallel, dplyr, ggcorrplot, ggmap, ggplot2, ggsn, ggthemes, glmnet, gstat, h2o, kableExtra, moments, osmdata, parallel, patchwork, raster, RColorBrewer, recommenderlab, rgdal, rgeos, runner, sf, showtext, sp, summarytools, tidyquant, tidyr, tidyverse, timechange, twilio and viridis.

**CRediT authorship contribution statement**

**Francisco Rowe:** Conceptualisation, Methodology, Software, Validation, Formal analysis, Investigation, Data Curation, Resources, Writing – Original Draft, Writing – Review & Editing, Visualisation, Supervision, Project administration. **Michael Mahony:** Software, Validation, Formal analysis, Data Curation, Writing – Original Draft, Writing – Review & Editing, Visualisation. **Sui Tao:** Conceptualisation, Data Curation, Writing – Original Draft, Writing – Review & Editing.

**Competing interests**
None

Supplementary Information
Machine Learning for Near-Real Time Bus Ridership Prediction During Extreme Weather

Francisco Rowe[1,*], Michael Mahony[1], Sui Tao[2]

[1]*Department of Geography and Planning, University of Liverpool, Liverpool, UK*
[2]*Faculty of Geographical Science, Beijing Normal University, Beijing, China*
*Corresponding author: F.Rowe-Gonzalez@liverpool.ac.uk

**Supplementary Table 1**. Summary of keys and values used to create urban amenity variables from OpenStreetMap

OSM categorises map features into keys and values. For example, the key of 'shop' contains a range of values, including 'general', 'electrical' and 'department_store'. From these features, 16 themes were selected to be used as variables. These are summarised in Section 3.4 and the table below provides a full description of the keys and values.

For each variable, the relevant features were called from OSM as spatial objects, such as points, lines, polygons, multilines and multipolygons. A 400m buffer was then constructed around every bus stop. This distance was chosen as it is frequently used by urban planners as an upper limit of how far people will walk to and from a bus stop (Foda and Osman 2010; Hess, 2012). A spatial index was then built using the st_intersects() function within the R package 'sf', to count how many spatial objects overlap with each 400m buffer. This resulted in a numerical estimate as to the density of different types of urban attractions within walking distance of each stop.

| Variable | Key | Values |
|---|---|---|
| Sustenance (Sn) | amenity | bar, bbq, biergarten, cafe, fast food, food court, ice cream, pub, restaurant, internet cafe, kitchen |
| Education (Ed) | amenity | college, driving school, kindergarten, language school, library, toy library, music school, school, childcare |
| University (U) | amenity | university |
| Transport (Tr) | amenity | bus station, ferry terminal, taxi |
| Financial Services (Fn) | amenity | atm, bank, bureau de change |
| Healthcare (Hc) | amenity | clinic, dentist, doctors, hospital, nursing home, pharmacy, social facility, veterinary |
| Small Day Ent (SE) | amenity | arts centre, community centre, fountain, social centre |
| Large Day Ent (LE) | amenity | cinema, planetarium, theatre, studio, dive centre |
| Night Ent (NE) | amenity | brothel, gambling, nightclub, swingerclub, stripclub |
| Religious (Rl) | amenity | monastery, place of worship |
| Civic (Cv) | amenity / office | courthouse, embassy, public building, townhall / diplomatic, government |
| Infrastructure (In) | amenity | conference centre, crematorium, grave yard, marketplace, police, post depot, post office, prison, public bath, recycling |
| Leisure (L) | leisure | adult gaming centre, firepit, fitness centre, sauna, sports centre, swimming pool |
| Shops (Sh) | shop | all values were used |
| Tourism (Tr) | tourism | hotel, picnic site |

**Supplementary Figure 1**. Variable importance plots by machine learning algorithm. Variable importance is determined by calculating the relative influence of each variable: whether that variable was selected to split on during the tree building process, and how much the squared error (over all trees) improved (decreased) as a result. Feature importance for each variable is scaled between 0 and 1, with scores closer to 1 denoting higher importance.

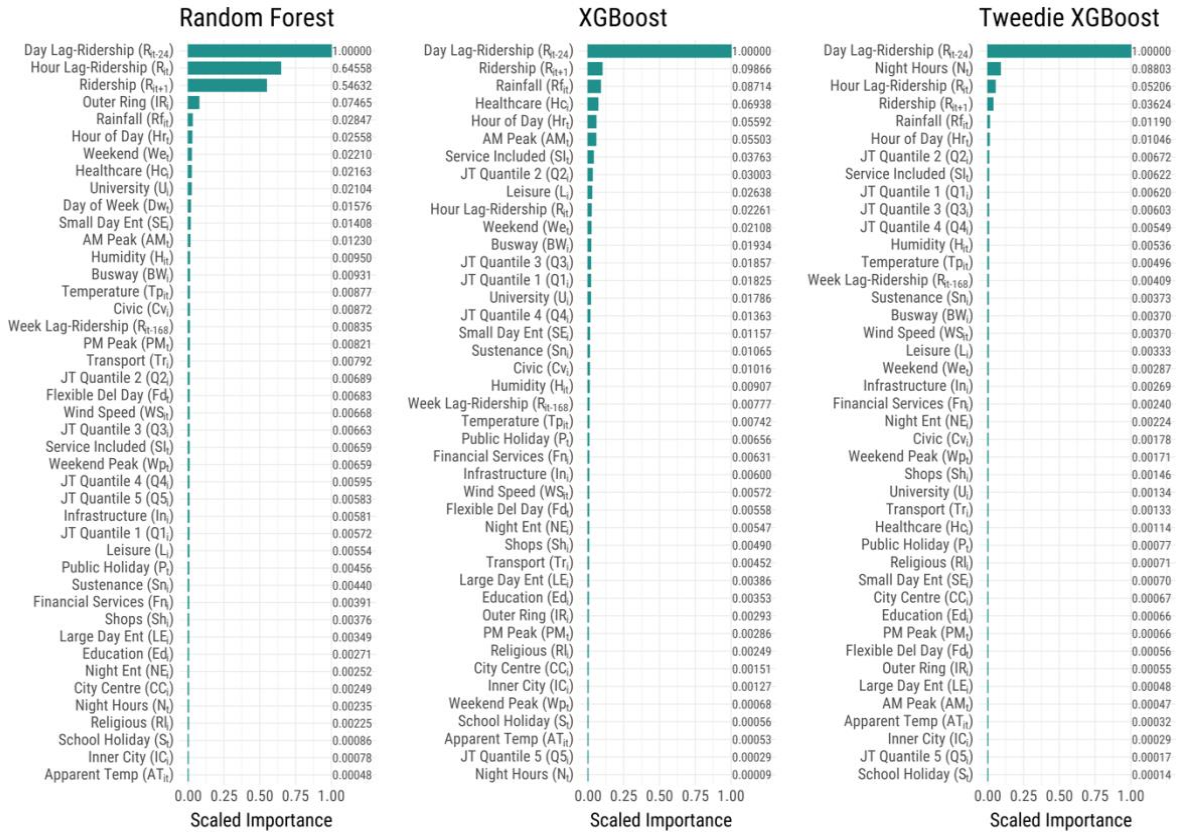